\documentclass[12pt]{iopart}
\usepackage{graphicx}
\usepackage{amssymb}
\usepackage{bm}
\begin{document}
\title[Size calibration due to dipole-monopole interaction]{
	Size calibration of strained epitaxial islands due to dipole-monopole interaction}
\author{V. I. Tokar and H. Dreyss\'e}
\address{IPCMS, Universit{\'e} de Strasbourg--CNRS, UMR 7504, 
23 rue du Loess, F-67034 Strasbourg, France}
\ead{tokar@unistra.fr}
\ead{hugues.dreysse@unistra.fr}
\begin{abstract}
Irreversible growth of strained epitaxial nanoislands has been studied
with the use of the kinetic Monte Carlo (KMC) technique. It has been
shown that the strain-inducing size misfit between the substrate and the
overlayer produces long range dipole-monopole (d-m) interaction between
the mobile adatoms and the islands.  To simplify the account of the long
range interactions in the KMC simulations, use has been made of a modified
square island model.  Analytic formula for the interaction between the
point surface monopole and the dipole forces has been derived and used to
obtain a simple expression for the interaction between the mobile adatom
and the rectangular island.  The d-m interaction was found to be longer
ranged than the conventional dipole-dipole potential. The narrowing of the
island size distributions (ISDs) observed in the simulations was shown
to be a consequence of a weaker repulsion of adatoms from small islands
than from large ones which led to the preferential growth of the former.
Furthermore, similarly to the unstrained case, the power-law behavior
of the average island size and of the island density on the coverage has
been found.  In contrast to the unstrained case, the value of the scaling
exponent was not universal but strongly dependent on the strength of
the long range interactions.  Qualitative agreement of the simulation
results with some previously unexplained behaviors of experimental ISDs
in the growth of semiconductor quantum dots was observed.\\

 \end{abstract}
\maketitle
\section{Introduction}
The phenomenon of self-assembly of size-calibrated coherent nanoislands
taking place in some heteroepitaxial systems during strained epitaxy
has been extensively studied for more than two decades because of its
prospective use in microelectronics \cite{rmp_dots,aqua_growth_2013}.
It seems to be well established that the phenomenon is governed
by the elastic strain in the overlayer caused by the lattice
size misfit with the substrate \cite{rmp_dots,aqua_growth_2013}.
Usually, it is assumed that the main role in the size calibration
(SC) play the long range forces propagated via the elastic strain
in the substrate \cite{lau_kohn,marchenko_parshin2,wires}.  However,
explicit growth simulations with the use of the kinetic Monte Carlo
(KMC) technique within the models accounting for such forces in
\cite{indirect_interactions3,aqua_frisch,aqua_interrupted_2013}, did not
show any narrowing of the island size distributions (ISDs).  Moreover,
in \cite{indirect_interactions3} even some broadening of the ISD was seen.
Notably, the broad ISDs obtained in the simulations were very similar to
those seen during irreversible growth \cite{2006}. So by all evidence the
growth in \cite{indirect_interactions3,aqua_frisch,aqua_interrupted_2013}
was controlled by kinetics; this is farther supported by the fact that in
thermodynamically controlled strained epitaxy in \cite{shchukin,JPhys13}
ISDs narrowing were observed.  These results seems to suggest that the
SC is implausible under conditions of kinetically controlled growth.
Such growth, however, usually takes place at smaller temperatures and
at faster deposition rates than the thermodynamically limited growth
\cite{shchukin} which presents some important practical advantages.
For example, smaller substrate-deposit interdiffusion and so better
control in heteroepitaxial growth \cite{ultrasmall}.  Therefore,
the question of whether the SC may be achieved under conditions of
kinetically controlled growth is of considerable practical interest.

The aim of the present paper is to suggest a new mechanism of the SC
during irreversible growth underlain by the repulsive long-range forces
induced by the misfit strain.  In contrast to the thermodynamically
controlled growth when the conventional dipole-dipole (d-d) interatomic
interactions \cite{lau_kohn,marchenko_parshin2} are sufficient to
ensure the SC \cite{shchukin}, the d-d forces are too weak to assure
SC in the kinetically controlled case, as the explicit simulations
in \cite{indirect_interactions3,aqua_frisch,aqua_interrupted_2013}
and our arguments in section \ref{forces} below show.  However, it is
known that besides the dipole forces at the strained surfaces there also
exist monopole forces that are of longer range than the d-d interactions
and in the case of the step-bounded surface structures, such as steps,
islands, and pits play a dominant role in their energetics and kinetics
\cite{wires,comment,tersoff_step-bunching_1995,jesson_morphological_1996,%
li_equilibrium_2000,q_platelets,bistability}.

In the present paper we will show that the monopole forces interacting
with the force dipoles induced by the mobile adatoms play similarly
important role in the growth kinetics. In particular, they provide a
mechanism of the SC during irreversible growth.  As will be shown in
section \ref{forces}, in the presence of d-m interactions the strength
of the repulsion between the island and the mobile atom may grow with
the island size in such a way that the atoms will avoid larger islands
by preferentially attaching to smaller ones, thus making the island
sizes more homogeneous.  The explicit confirmation of this mechanism
in explicit KMC simulations will be provided in section \ref{KMC}.
But because the main difficulty in studying the strained epitaxy with
the KMC technique is the necessity of accounting for the long-range
interactions acting between all atoms in the simulated system,
essential simplifications and approximations are necessary to make the
simulations feasible.  The approximations used in KMC simulations in
\cite{shchukin,indirect_interactions3,aqua_frisch,aqua_interrupted_2013}
with the d-d interactions are not suitable for the d-m case.
Therefore, in sections \ref{islands} we will introduce a simple
model of strained islands, in section \ref{forces} will calculate
the strain-induced interactions between the adatoms and the islands
in the planar approximation,  and in section \ref{KMC} will explain
the approximate description of the capture of the mobile adatoms
by the square islands.  Also, in this section we discuss the
applicability of our results to the explanation of experimental
data on the SC of the quantum dots in Ge/Si(001) system studied in
\cite{chaparro_evolution_2000,drucker_diffusional_1997}.

In the final section we present our conclusions.
\section{\label{islands}The model of strained islands}
The main problem in KMC simulation of many-body systems with long-range
interactions is that to simulate the change of the position of a single
atom, the energy of its interaction with all other atoms in the system
need be calculated first.  And in the case of the Metropolis algorithm
\cite{KMCreview} which is quite appropriate in this case (at sufficiently
high temperatures, at least), the calculation need to be performed twice:
for the initial and the final positions of the moving atom. And in
the end the move can be discarded by the algorithm.

Because one is usually interested in the thermodynamic limit,
the simulated system should be reasonably large to mitigate the
finite-size effects.  For example, in our simulations we used, following
\cite{shchukin}, the system consisting of $250\times250$ sites on a
square substrate lattice.  With maximum coverage $\theta=0.2$ chosen to
remain in precoalescence growth regime \cite{2006} our system contained
up to 12500 atoms.  In order to be able to calculate the interaction
with all of them at each KMC step we had to simplify the task by
adopting several simplifying assumptions.  In particular, we assumed
that due to the fast intraisland diffusion the islands acquire simple
quasiequilibrium symmetric shapes that can be found with the use of the
Wulff construction \cite{daruka_dislocation-free_1997}.  This approach
has been widely used in simulations of unstrained epitaxial growth
\cite{bartelt_nucleation_1993,bales_chrzan,indirect_interactions3,%
li_evans03,defects_ini,approx_coincid} so below we adopt it to
our needs.  In particular, this will allow us to make comparison
of our results with experiments on the growth of three-dimensional
quantum dots (QDs) because crucial to our SC mechanism will be only
the island-substrate interface and the shear strain propagated by it.
So though in our simulations we use monolayer-high islands, if the
base layer of a QD is size calibrated, the height (hence, the volume)
that can be found via the Wulff construction will be also subject to the
SC \cite{daruka_dislocation-free_1997}.  In the present paper, however,
we will restrict our simulations to the simplest case of monolayer-high
(or submonolayer \cite{sml_qds13}) islands.

Submonolayer islands on the square substrate lattice at low temperature
will strive to acquire rectangular shapes \cite{molphys}.  In the harmonic
approximation the elastic forces in orthogonal directions decouple.
This allows one to treat the elastic relaxation independently within
each linear atomic chain that compose the island \cite{PRB,JPhys13}.
Another simplification is to neglect the displacements of the surface
atoms in the direction perpendicular to the surface, as suggested in
\cite{lau_kohn,marchenko_parshin2,steinfort_strain_1996}.  This leaves
us with one-dimensional (1D) chains of atoms lying at the rigid substrate
\cite{PRB,JPhys13}.  Their relaxation can be described within the harmonic
Frenkel-Kontorova model as follows \cite{PRB}.  In this model the atoms
are harmonically bound to the substrate sites, so the point monopole
forces applied to the deposition cites in the in-plane directions 
are proportional to the atomic displacements and the stiffness $k$.

\begin{figure}[h]
	\centering
	\includegraphics[scale=0.75]{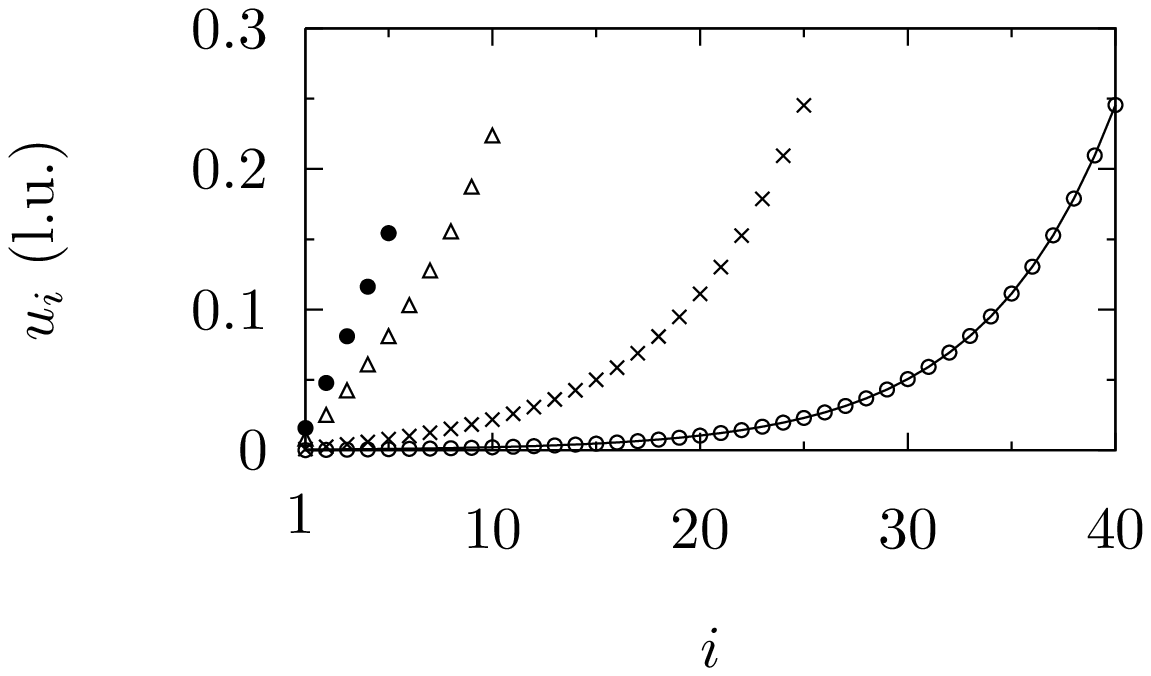}
	\caption{Symbols: Atomic displacements calculated according to
	(\ref{u_j}) with $\alpha=0.025$ and $f=0.042$ for the chains
	consisting of (from left to right) 10, 20, 50 and 80 atoms. Only
	one half of the displacements are shown; the other half has the
	same values but the opposite sign.  The solid line shows the
	displacements of the end 40 atoms of an infinitely long chain.}
	\label{fig:fig5}
\end{figure}

The displacements $u_j$ of atoms $j=1,2,\dots,l$ within a chain consisting
of $l$ atoms can be calculated as
\begin{equation}
	u_j = f \sinh[\phi (2 j-l-1)]/[\sqrt{\alpha}\cosh(\phi l)],
	\label{u_j}
\end{equation}
where $f$ is the size misfit between the substrate and the overlayer,
$\alpha=k/k_{NN}$ the ratio of the rigidities of the elastic springs binding
the atom to the substrate $k$ and to the nearest neighbor atom $k_{NN}$ and 
\begin{equation}
	\phi=\ln(\sqrt{1+\alpha/4}+\sqrt{\alpha}/2).  
	\label{phi}
\end{equation}
In (\ref{u_j}) and everywhere below all lengths are measured in the
substrate lattice units (l.u.). So that the numerical value of the misfit
parameter $f$ coincides with the relative misfit.  In Ge/Si(001), for
example, $f=0.042$ because the relative misfit in the system is 4.2\%.
We used this value of $f$ together with small value of $\alpha$ in figure
\ref{fig:fig5} to schematically illustrate the distribution of the atomic
displacements inside atomic chains of different length.

The small value of $\alpha$ in the figure was chosen to be close to the rigid-core
case corresponding to $k_{NN}\to\infty$ and $\alpha\to0$.  As can be seen from
(\ref{u_j}) and (\ref{phi}), in this case 
\begin{equation}
	u_j=f(2j-l-1)/2,
	\label{k_infty}
\end{equation}
i.\ e., the atomic displacements depend linearly on the distance from
the middle of the chain \cite{JPhys13}.  As we will show in the next
section, the SC may take place in islands where the atomic displacements
and as a consequence the monopole forces grow with the island size.
It is this type of islands that we will use in our KMC simulations in
section \ref{KMC}.  But obviously that in reality the atoms are not
absolutely rigid and the compressibility ($1/k_{NN}$) is finite, though
it can be small.  From figure \ref{fig:fig5} it is seen that the displacements
follow the linear behavior (\ref{k_infty}) only in sufficiently
small islands.  In large islands the displacements and the
monopole forces saturate and phenomenologically can be described
by constant monopole force density at the island edge, as in
\cite{wires,comment,tersoff_step-bunching_1995,jesson_morphological_1996,%
li_equilibrium_2000,q_platelets,bistability}.  In this case the small
and large islands repulse mobile monomers with similar force so our
SC mechanism became inefficient.  

The discrete strain can be calculated as (assuming the chain is oriented
in the $x$ direction)
\begin{equation}
	\varepsilon_{xx}=\Delta u/\Delta x = u_{i+1}-u_i
	\label{epsilon}
\end{equation}
because the $x$-coordinate difference is equal to 1 l.u..  As can be
seen from figure \ref{fig:fig5}, the average strain in small islands
has appreciable value which diminishes with growing island size.
In large islands strain remains only in the ends of the chain, so the
average strain will tend to zero as $1/l$.  It is exactly the behavior
found in square Co islands on the Cu(001) surface in {\em ab initio}
molecular dynamics simulations in \cite{stepanyuk_in_plain_strain}.
The authors assess that the saturation starts in islands of $\sim2$~nm size.
Size-calibrated metallic islands of similar sizes are needed to produce
efficient catalysts \cite{catalytic_Au}, so our SC mechanism can be of
practical importance in this field.
\section{\label{forces}Long range elastic interactions on the surface}
In this section we derive expressions for the substrate-propagated elastic
interactions to be used in KMC simulations in section \ref{KMC}.  Due to
the long-range nature of the interactions \cite{ll7}, the hopping adatom
interacts with all other atoms on the surface.  In order to calculate
the interaction with the large number of atoms at each atomic hop,
computationally efficient expressions for the interaction energy are
needed.

As was explained previously, such expressions can be derived
for epitaxial islands of simple geometry.  Thus, following
\cite{bartelt_nucleation_1993,bales_chrzan,indirect_interactions3,li_evans03,%
defects_ini,approx_coincid} we assume that at low temperature the
islands on the square substrate lattice acquire rectangular shapes and
will derive the expressions for the interaction of the mobile monomers
with such islands.
\subsection{The dipole-dipole interaction}
First we show that the conventional dipole forces are not sufficiently
strong for the above SC mechanism to be operative.  The potential
acting between two adatoms at distance $r$ apart in this case is
\cite{lau_kohn,marchenko_parshin2} 
\begin{equation}
	V^{\rm d-d}=\frac{\gamma}{r^3}, \label{V_at}
\end{equation} 
where $\gamma>0$ is the interaction strength and $r$ , so $r$ is
dimensionless and the closest atoms are 1 l.u.\ apart.

Now assuming the atoms form a rectangle with sides $a$ and $b$ along
the $x$ and $y$ directions, respectively, and summing contributions of
the form of (\ref{V_at}) over all atoms within the island in the
continuum approximation one gets:
\begin{equation}
	V_{ab}^{\rm d-d}(x,y)={\gamma}\bigg[\bigg[\frac{\bar{r}}{\bar{x}\bar{y}}
		\bigg]_{\bar{x}_-}^{\bar{x}_+}\bigg]_{\bar{y}_-}^{\bar{y}_+}.
	\label{V_isl0}
\end{equation}
Here two-dimensional radius vector $\mathbf{r}=(x,y)$ points from the
island center which we placed for simplicity at the coordinate origin $(0,0)$
to the position of the mobile atom; $\bar{x}_{\pm}=x\pm a/2$,
$\bar{y}_{\pm}=x\pm b/2$, and $\bar{r}=(\bar{x}^2+\bar{y}^2)^{1/2}$.
The square brackets in  (\ref{V_isl0}) denote the substitution of
four possible combinations of $\bar{x}_{\pm},\bar{y}_{\pm}$ for $\bar{x}$
and $\bar{y}$ with necessary signs.

Rectangular islands were grown in the KMC simulations in
\cite{indirect_interactions3} but no narrowing of the ISDs was found.
Equation (\ref{V_isl0}) allows us to understand this.  According to
\cite{tokar_rigorous_2013} the island capture numbers in the rate
equations that define the rate at which the islands capture the mobile
adatoms is proportional to the density of the adatoms at the sites
situated one hopping step away from the island.  They may be called
the island nearest neighbors (NN).  At high values of the diffusion to
deposition rates ratio that we are going to simulate in section \ref{KMC}
the growth is in the thermodynamically controlled regime the distribution
of mobile atoms is very close to equilibrium \cite{shchukin,JPhys13}
so their density can be assessed as
\begin{equation}
	N_{1NN} \propto \exp[-E(i_{NN})/k_BT],
	\label{N_1NN}
\end{equation} 
where $E(i_{NN})$ is the energy of the adatom at some point $i_{NN}$ NN to
the island.  Let us for definiteness consider the point $i_{NN}=(a/2+1,0)$,
at 1 l.u.\ distance from the middle of the edge of length $a$ of an island
centered at the coordinate origin $(0,0)$.  From (\ref{V_isl0}) it is
easy to see that at large $a,b$ $E(i_{NN})=V_{ab}^{\rm d-d}(a/2+1,0)$
saturates to a constant value.  Thus, the mobile atoms can reach both
large and small islands with equal ease, so the d-d repulsion does not
cause SC during irreversible growth.
\subsection{The monopole-dipole interaction} 
The dipole-dipole interactions correspond to the situation when each
adatom is positioned in the geometric center of the substrate lattice
unit cell so all atomic displacements in the substrate are also symmetric
and the resulting force distribution corresponds to the force dipole.
Such a behavior would describe rather non-strained situation because the
presence of a positive misfit $f$ means that the adatoms are too big
to fit into the substrate cells without pushing their neighbors. This
situation can be described within the model of the rigid-core adatoms
\cite{JPhys13} which should be adequate for situations where $f>0$ and
the atomic relaxation in the island base layer is small.  Thus, it is
easy to see from (\ref{k_infty}) that if the diameter of the rigid core
is $1+f$, then the atom with the coordinates $\mathbf{r}=(x,y)$ inside the
island centered at the origin will be displaced by its neighbors from the
center of the lattice cell it belongs to in the direction $\mathbf{r}$ as
\begin{equation}
	\Delta\mathbf{r}=f\mathbf{r}.
	\label{delta_r}
\end{equation}
\cite{JPhys13}.
But the adatoms are bound to the substrate cell centers by an effective
harmonic spring with the spring constant $k$.  So the displaced
adatom at position $\mathbf{r}$ within the island will exert on the
substrate a shear force in the $x-y$ plane with the density
\begin{equation}
	\mathbf{F}^m(\mathbf{r}^{\prime})=k\Delta 
	\mathbf{r}\delta(\mathbf{r} -\mathbf{r}^{\prime})
	=kf\mathbf{r}\delta(\mathbf{r} -\mathbf{r}^{\prime}),
	\label{F_m}
\end{equation}
where for simplicity we again resorted to the continuum approximation.
We supplied the force density with the superscript ``$m$'' to
stress its monopole character.  Of course, there always exists an
atom at $\mathbf{r}=-\mathbf{r}$ with the opposite force density
so that at large distances from the island the potential will have
the dipole-dipole asymptotic of the type of  (\ref{V_at}).
In large islands, however, the distance $2r$ between the atoms
can be large so an adatom approaching the island boundary will
experience effectively the monopole forces.  

The energy of interaction of an adatom with a force monopole is the work
performed by the dipole force distribution ${\mathbf F}^d$ along the
field of displacements ${\mathbf u}$: 
\begin{equation}
	V^{\rm d-m}(\mathbf{r})=-\int\mathbf{F}^d({\mathbf{r-r}^{\prime}})\cdot
	\mathbf{u}(\mathbf{r}^{\prime})\,dx^{\prime}dy^{\prime}.  
\label{Vdm0}
\end{equation}
The dipole force distribution produced by an adatom at ${\mathbf
r}$  is \cite{marchenko_parshin2}
\begin{equation}
	{\mathbf F}^d({\mathbf{r-r}^{\prime}})=A\mathbf{\nabla} 
	\delta({\mathbf{r-r}^{\prime}}),
	\label{Fd}
\end{equation}
where $A$ is a constant.  The displacement field due to the monopole
force $\mathbf{F}$ applied at the coordinate origin for the isotropic
case was calculated in \cite{ll7}.  So substituting expressions (8.19)
for $\mathbf{u}(\mathbf{r})$ from that reference together with (\ref{Fd})
into (\ref{Vdm0}) one finds after some algebra
\begin{equation}
	V^{\rm d-m}(\mathbf{r})=\frac{A(1-\nu)}{2\pi\mu}\frac{\mathbf{F}\cdot 
	\mathbf{r}}{r^3},
	\label{d_m}
\end{equation}
where $\nu$ is the Poisson ratio and $\mu$ the shear modulus.
As is seen, the interaction in  (\ref{d_m}) is longer-ranged than
in (\ref{V_at}).  We remind that everywhere in the present study we,
following \cite{lau_kohn,marchenko_parshin2}, consider only 2D in-plane
forces and other vectors, so the component of the monopole force in the
direction perpendicular to the surface was set to zero in expressions
(8.19) from \cite{ll7}.  In more sophisticated models of strain in
epitaxial islands (see, e.\ g., \cite{stepanyuk_orthogonal_strain})
this component is non-vanishing and should be included in (\ref{d_m})
along the line of derivation presented above.

Now substituting  (\ref{F_m}) into  (\ref{d_m}) and integrating
over the rectangular island $a\times b$ centered at the origin 
$(0,0)$ one gets the potential of interaction with the adatom 
placed at the point $(x,y)$ external to the island as
\begin{equation}
	V_{ab}^{\rm d-m}(x,y)=C\bigg[\bigg[(x+\bar{x})\ln({\bar{y}+\bar{r}})
		    +(y+\bar{y})\ln({\bar{x}+\bar{r}})
                       \bigg]_{\bar{x}_-}^{\bar{x}_+}\bigg]_{\bar{y}_-}^{\bar{y}_+},
	\label{V_isl}
\end{equation}
where $C={Akf(1-\nu)}/{2\pi\mu}$ and other notation is the same as in 
(\ref{V_isl0}).

To farther simplify the simulations, below  we
will assume that the islands are of square shape
\cite{bartelt_nucleation_1993,bales_chrzan,li_evans03} because in the
case of weak elastic forces and small islands we are going to study the
aspect ratios of the rectangular islands are known to be close to unity
\cite{indirect_interactions3,bistability}.

The potential (\ref{V_isl}) on the nearest-neighbor distance from the square island boundary
\begin{equation}
	E(i_{NN})=V_{aa}^{\rm d-m}(a/2+1,0)|_{a\to\infty}\sim C a\ln a 
	\label{asymptotics}
\end{equation}
which means that in contrast to the potential (\ref{V_isl0}) derived from
the dipole-dipole interaction, the potential based on the dipole-monopole
forces does differentiate between mobile adatom capture by large and
small islands, as can be seen from (\ref{N_1NN}),---thus providing a
mechanism for kinetically controlled SC.
\section{\label{KMC}KMC simulations of the growth of the square islands}
To assess the efficiency of the proposed mechanism, kinetic
Monte Carlo (KMC) simulations were carried out with the
use of a variant of the square-island model developed in
\cite{bartelt_nucleation_1993,bales_chrzan,li_evans03}.  As explained
in the previous section, to simplify the calculation of the elastic
forces we slightly modified the model by applying the approach of
\cite{defects_ini,approx_coincid} to the square islands instead of the
circular ones.  Namely, we assumed that the side length of the island
$a=\sqrt{s}$, where $s$ is the island size.  Then the square capture
zone \cite{2006} with the side length $a+2$ is formed by surrounding the
island with a strip of width 1.  Any atom that enters the capture zone
either by direct impingement or via the hopping diffusion is irreversibly
caught by the island whose size becomes $s+1$.
\subsection{Growth in the absence of elastic interactions}
To validate our KMC setup we first carried out the simulations without the
long-range interactions.  Two points could arose concern in connection
with our approach. First, because the capture of the monomers by
the islands is different from the conventional square island model
\cite{bartelt_nucleation_1993,bales_chrzan,li_evans03},  the question
arises on whether the physics of the growth remains qualitatively 
the same.  Second,
because of the difficulties with accounting for the interaction of
the diffusing monomer with all atoms in the system, the size of the
simulated lattice was chosen to be $250\times250$ sites which is smaller
than typically used for the simulations of the growth without long-range
interactions.  Because our main interest in the present study are the ISDs,
we compared the total number of atoms in two-dimensional square islands
obtained in our approach with corresponding results from 
\cite{indirect_interactions3,li_evans03}.  The diffusion constant was calculated
according to the standard expression 
\begin{equation*}
	D=\nu_{\rm att}\exp(-E_d/k_BT)
\end{equation*}
with typical values for the attempt frequency $\nu_{\rm att}=1$~THz, the
diffusion barrier $E_d=0.7$~eV \cite{JPhys13}, and the deposition rate
1.4~ML/min \cite{chaparro_evolution_2000}.  To gather good statistics
simulations were repeated from 160 to 480 times so that the statistical
errors in our data are very small usually not exceeding the sizes of
the symbols used to plot the data.

\begin{figure}
	\begin{center}
		\includegraphics[scale=0.5]{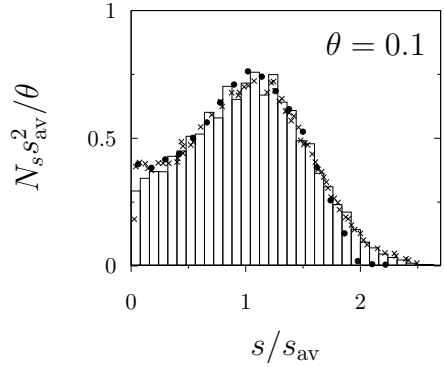}
	\end{center}
	\caption{Simulated ISD for irreversible growth at 400 $^\circ$C
	in the absence of misfit strain in our model (histogram); for comparison
	are shown simulation data from \cite{indirect_interactions3}
	(filled circles) and  \cite{li_evans03} (crosses)}
	\label{fig:isd}
\end{figure}

\begin{figure}
	\begin{center}
		\includegraphics[scale=0.5]{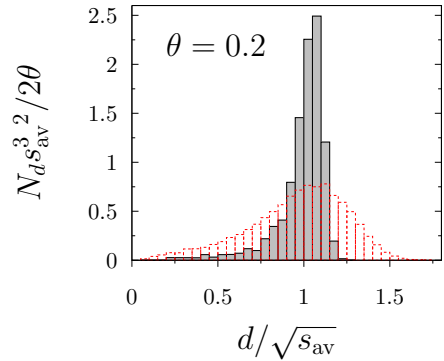}
	\end{center}
	\caption{Shaded histogram---distribution of the diameters (defined
	as $d=\sqrt{s}$) of the islands grown in the KMC simulations
	at 400~$^\circ$C with the elastic interaction corresponding to
	$\gamma=20$~meV. The dashed histogram is the diameter distribution
	for $\gamma=0$.}
	\label{fig:dsd}
\end{figure}

We first checked the soundness of our approximations by simulating
the growth without elastic forces and comparing the results with
simulations on similar models of compact islands.  The data shown in
figure \ref{fig:isd} are plotted in the scaling variables \cite{2006},
as is conventional in the precoalescence regime at coverage $\theta$
not exceeding $\sim0.2$:
\begin{equation}
	N_s=\theta f(s/s_{\rm{av}})/s_{\rm{av}}^2,
	\label{scaling}
\end{equation}
where $N_s$ is the density of islands of size $s$, $s_{\rm{av}}$ is the
average island size, and $f$ a universal scaling function.
As is seen, the agreement is very good; most importantly, no ISD narrowing
is seen in our data which means that the SC obtained in the simulations
shown on figure \ref{fig:dsd} is due to the strain and not because of
the approximations made.
\subsection{KMC simulation of strained epitaxy}
To simulate the growth with realistic strain in our model we need to
chose the value of the constant $C$ in (\ref{V_isl}).  For consistency
we chose it in such a way that islands with $a=1$ corresponding to
isolated adatoms asymptotically reproduced the dipole-dipole potential
(\ref{V_at}).  After some algebra the asymptotic of (\ref{V_isl}) was
found to be $C/(12r^3)$, so $C=12\gamma$ would assure that at distances
larger than 1 l.u.\ the interaction of an island consisting and of an
adatom will have the same strength as the adatom-adatom interaction
(\ref{V_at}).

The range of numerical values of $\gamma$ used in the
simulations was chosen according to estimates made in
\cite{aqua_frisch,indirect_interactions3,lau_kohn}.  Because the
estimated values vary in a rather broad range and {\em ab initio}
estimates of non-local interactions are known to be unreliable
\cite{nucl_impurities+abinit}, the simulations were carried out for five
values of $\gamma=0$, 2.5, 5, 10 and 20 meV.

\begin{figure}
	\begin{center}
		\includegraphics[scale=0.5]{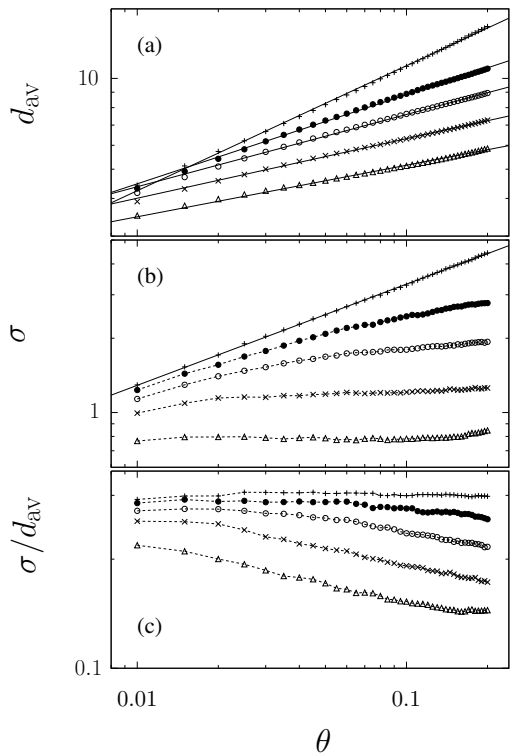}
	\end{center}
	\caption{Log--log plots of the coverage dependences of the mean
	islands diameters (a), of the diameter distributions
	dispersions (b), and of the dispersion to the diameter
	ratios (c) for five values of the interaction parameter $\gamma$:
	0 (+), 2.5~meV ($\bullet$), 5~meV ($\circ$), 10~meV ($\times$),
	and 20~meV ($\Delta$). Solid lines are linear fit to the data;
	dashed lines are guides to the eye. Note that only $\gamma=0$
        data exhibit the scaling behavior.}
	\label{fig:3figs}
\end{figure}

\begin{figure}
	\begin{center}
		\includegraphics[scale=0.5]{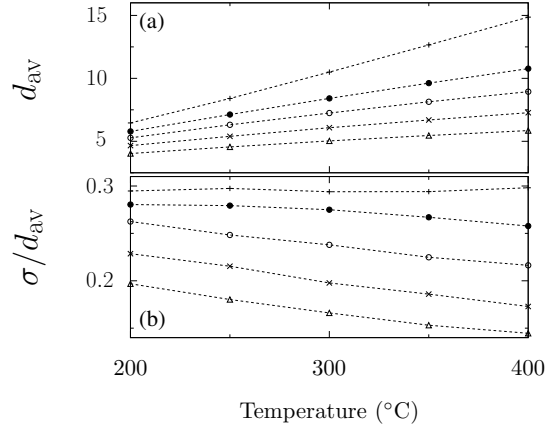}
	\end{center}
	\caption{Temperature dependences of the mean island diameter 
		$d_{\rm av}$ (a) and of the ratio of the diameter 
		distribution dispersion $\sigma$ to $d_{\rm av}$ (b)
		at coverage $\theta=0.2$ for different strength of the
		elastic interaction; notation is the same as in figure 
		\ref{fig:3figs}.}
	\label{fig:2figs}
\end{figure}

The main results of our KMC 
simulations are shown in figures \ref{fig:3figs} and \ref{fig:2figs} 
Two conclusions can be drawn from the data.  The value of index $\omega$ 
introduced in \cite{1992} from the power-law dependence
\begin{equation}
	s_{\rm av}\propto \theta^{-\omega}
	\label{omega}
\end{equation}
can be found from our data on $d_{\rm av}\simeq\sqrt{s_{\rm av}}\propto
\theta^{-\omega/2}$.  This index is convenient for experimental
measurement because it is directly connected to the index characterizing
the total island density
\begin{equation}
	N=\theta/s_{\rm av}\propto\theta^{1+\omega}
	\label{N}
\end{equation}
Our first conclusion is that the index strongly depends on the strength
of the elastic forces, especially at small $\gamma$, as can be seen from
table \ref{tab:table1}.
\begin{table}
	\centering
	\begin{tabular}{c|l}
		{$\gamma$}{(meV)}& \hspace{1.25em}$\omega$\\ \hline
0      &  -0.82\\
2.5 &  -0.59\\
5.0  & -0.48 \\
10.   & -0.39\\
20.   & -0.34
	\end{tabular}
	\caption{Dependence of the scaling exponent $\omega$ as defined in
		(\ref{omega}-\ref{N}) on the strength of the long-range
		interaction parameter $\gamma$ (\ref{V_at}) }
	\label{tab:table1}
\end{table}
Experimentally the smaller values of $\omega$ may look as the growth
saturation, as seems to be the case in \cite{bimodal0}.  The second
conclusion is that in the presence of strain the dispersion $\sigma$ of
the island diameter distribution (IDD) exhibits a saturated behavior,
as can be seen from figure \ref{fig:3figs}(b).  In combination with
monotonous growth of $d_{\rm av}$  the dispersion to mean diameter ratio
diminishes with coverage in qualitative agreement with experimental
data \cite{bimodal0}.
\subsection{\label{discussion}Discussion}
The dispersion to mean diameter ratios in figures \ref{fig:3figs}(c)
and \ref{fig:2figs}(b) are not as small as in some experimental data
that exhibit the best cases of SC.  One of reasons is that in our
calculations we used the standard formulas of statistics and took into
account all available IDD data from the smallest to the largest island
sizes.  Because they contain a tail in the IDD curves at small diameters
(see figure \ref{fig:dsd}), the values of $d_{\rm av}$ are smaller than
diameter values at the IDD maxima.  This augments both $\sigma$ and
$\sigma/d_{\rm av}$ values. Here it is pertinent to note that similarly
asymmetric IDDs are rather commonly observed experimentally (see, e.\
g., \cite{bimodal0,drucker_diffusional_1997,chaparro_evolution_2000,%
bimodal_annealing_2010}).  But because experimental data as a rule
contain more different structures than exists in our simple model,
size distributions for islands of different morphologies are usually
studied separately.  Therefore, usually only the data in the vicinity
of the maximum density of islands of a given morphology are taken into
account in the processing of experimental data.  Thus, for example, the
value of the diameter at the IDD maximum for a given kind of islands is
taken for $d_{\rm av}$ and the IDDs widths are calculated only in its
vicinity \cite{bimodal0,drucker_diffusional_1997,chaparro_evolution_2000}.
In \cite{bimodal0}, for example, the relative FWHM with respect to
$d_{\rm av}$ taken to be equal to $d$ at the IDD maximum was found to be
quite small (15\%).  But as can be seen from our figure \ref{fig:dsd},
in our case it has similar value $\sim14\%$.  It is also common to
fit experimental IDD with the Gaussian curve.  But for the latter
the WFHM$\approx2.35\sigma$ so the effective value of $(\sigma/d_{\rm
av})_{eff}$ with such data processing can be as small as 0.06.  Thus,
the SC mechanism proposed may underly even the best cases of SC seen
experimentally.
\section{Conclusion}
To conclude, in this paper we suggested a kinetic mechanism of the SC
in strained epitaxy. It differs from the kinetic mechanisms in the
presence of the Ostwald ripening \cite{bimodal0,bimodal_annealing_2010}
in that there is no need for atomic detachments.  This means
that the mechanism can be operative at smaller temperatures which
may have important practical advantages.  But because of similar
growth behavior, the mechanism can also contribute to the phenomena
attributed to the ripening.  Furthermore, it may underly the unusual
narrowing of IDD with temperature observed in the QDs growth in
\cite{drucker_diffusional_1997,chaparro_evolution_2000} (cf.\ our
figure \ref{fig:2figs}).  Such behavior is qualitatively different from
that observed both in thermodynamically controlled growth and in the
kinetically controlled growth in the absence of strain.

The proposed mechanism of SC heavily relies on the island size dependent
monopole forces due to the misfit shear strain in the substrate, so the
strength and spatial extent of the forces are of crucial importance.
The simple rigid-core model \cite{JPhys13} we used to illustrate our
mechanism is presumably good for small islands simulated in the present
paper and for those grown experimentally in \cite{ultrasmall}.  For very
large islands, however, saturation toward the constant monopole density
similar to that on the surface steps should be expected.  Calculations in
\cite{sigma_xz_1995} revealed the shear strain that linearly varies across
the QD/substrate interface in capped InAs/GaAs pyramids of at least 12 nm
in diameter.  Experimentally large interface shear strain of considerable
spatial extent caused by Ge/Si(001) QDs of an order of magnitude larger
diameter was observed in \cite{kohmura_controlling_2013}.  This may
mean that the mechanism described in the present paper contributes
to the SC of QDs of all sizes.  But even restricted to islands a few
nanometer in diameter, the mechanism would still be of considerable
practical interest.  The maximal catalytic efficiency is achieved in
size calibrated metallic islands of small diameters \cite{catalytic_Au}.
In semiconductor heteroepitaxy the quantum size effect that allows for
variation of the QD photoluminescence wavelength is operative only in
small QDs.  Finally, the model of monolayer-high islands that we studied
in the present paper can be taken as a starting point for modeling the
growth of the submonolayer QDs \cite{sml_qds13}.  Their small spatial
dimensions may allow for the fabrication of the most compact QD devices
due to dense QD packing.

Finally, in the course of our study we derived a simple expression for
the dipole-monopole interaction for in-plane displacements and forces
that can be generalized to 3D in case of necessity.  This expression
can be used in other studies of nucleation and growth unrelated to SC.
For example, in studying the influence on the growth of the surface steps,
islands and pits that are usually present in strained epilayers.

\end{document}